\documentclass{ws-ijmpd}
\usepackage{amsmath}
\usepackage{amssymb}

\begin{document}

\markboth{S.C. Ulhoa and E. P. Spaniol} { On the Gravitational
Energy-Momentum Vector in f(T) Theories}

%
\catchline{}{}{}{}{}
%

\title{On the Gravitational Energy-Momentum Vector in f(T) Theories }

\author{S. C. Ulhoa }

\address{Instituto de F\'{i}sica, Universidade de Bras\'{i}lia, 70910-900, Bras\'{i}lia, DF, Brazil.\\ Faculdade Gama, Universidade de Bras\'{i}lia, Setor Leste (Gama), 72444-240, Bras\'{i}lia, DF, Brazil.
\\
sc.ulhoa@gmail.com}

\author{E. P. Spaniol}

\address{
UDF Centro Universit\'ario and Centro Universit\'ario IESB (Campus Sul – Edson Machado), Bras\'ilia, DF, Brazil.\\
spaniol.ep@gmail.com }

\maketitle
\begin{history}
\received{\today} \revised{Day Month Year}
\end{history}

\begin{abstract}
This work is devoted to present and analyze an expression for the
gravitational energy-momentum vector in the context of f(T) theories
through field equations. Such theories are the analogous counterpart
of the well known f(R) theories, except using torsion instead
of curvature. We obtain a general expression for the gravitational
energy-momentum vector in this framework. Using the hypothesis of
the isotropy of spacetime, we find the gravitational energy for a
closed Universe, since construction of real tetrads that do not constrain the functional form of the Lagrangian density was not possible for an open Universe. Thus we find a vanishing gravitational energy
for the tetrad that we have used. 
\end{abstract}

\section{Introduction}
\noindent

Teleparallel gravity and general relativity are two different theories with the same field equations. In general relativity it is not possible to construct an expression for the gravitational energy in terms of the metric tensor and its second derivatives which give rise to the approach of energy-momentum pseudo-tensors. On the other hand in TEGR there is an expression for the gravitational energy-momentum vector, tested throughout the years, in terms of the tetrad field which is the dynamical variable of the theory. This feature will be explored in this article.

The Hilbert-Einstein lagrangian density, which gives the dynamics of
general relativity, can be generalized as a function of the Ricci
scalar. This establishes a wide class of lagrangian densities that
lead to what is known as f(R) theories~\cite{Felice}. Several
cosmological observations, such as supernovae
\cite{S.Perlmutter,1538-3881-116-3-1009}, cosmic microwave
background \cite{D.N.Spergel,0067-0049-170-2-377}, large-scale
structure \cite{M.Tegmark,PhysRevD.71.103515} and baryon acoustic
oscillations \cite{D.J.Eisenstein}, point to an exotic kind of energy
which is known as dark energy. The f(R) theories have been used
successfully to explain such observations. They can explain the anomalous
experimental data since it is added extra terms in the energy-momentum
tensor of matter fields in the Einstein equations, working as a
different source of energy density.

Such a success leads us to study the analogous generalization of the
lagrangian density of teleparallel gravity, known as f(T)
theories. Recently, models based on f(T) gravity were presented as
an alternative to inflationary models~\cite{ferraro}. Moreover, in
the literature we find some works that attempt to explain the
late-time accelerated expansion of the Universe in the context of
f(T) theories~\cite{Bengochea,Linder}.

Recently the energy of the Universe was obtained in the context of
teleparallel gravity~\cite{Ulhoa:2010wv}. Such a result was derived
considering the Universe as described by the FRW metric as well as with the use
of regularized expressions. This, in cosmology, is necessary to describe reference frames since it is possible to have
remanence of torsion in the flat space-time. Thus we intend to find
a general expression of the energy-momentum vector in the context of
f(T) theories, contained in an arbitrary volume V of the
three-dimensional spacelike hypersurface, $P^a$. The definition of
$P^a$ proved to be invariant under coordinate transformations and
transforms like a vector under global Lorentz transformations,
features that are essential to a true energy-momentum vector. Then,
as an application we obtain the gravitational energy of the universe
taking into account a well known tetrad obtained in the literature. Since the
theory is not invariant under local Lorentz transformations, choosing the best tetrad field is not a trivial task.

The paper is organized as follows: in section \ref{tel} the $f(T)$
gravity is presented, we obtain through the field equations an expression
for the gravitational energy-momentum vector. In section \ref{frw}
we apply our expression in the context of a homogeneous and
isotropic Universe. We obtain a vanishing gravitational energy for
all possible models, concerning the functional form of $f(T)$. We
have used a ``good'' tetrad which is a solution of the field
equations that does not constrain the functional form of the
lagrangian density. Finally in the last section we present some
concluding remarks.

\bigskip
Notation: space-time indices $\mu, \nu, ...$ and SO(3,1) indices $a,
b, ...$ run from 0 to 3. Time and space indices are indicated
according to $\mu=0,i,\;\;a=(0),(i)$. The tetrad field is denoted by
$e^a\,_\mu$ and the determinant of the tetrad field is represented
by $e=\det(e^a\,_\mu)$.\par
\bigskip

\section{Teleparallel Gravity and f(T) Theories}\label{tel}
\noindent

In this section we would like to establish the field equations of f(T) theories as well as a definition of energy-momentum vector.
However it is first necessary to recall some ideas of TEGR, which is an alternative theory of
gravitation, that is entirely equivalent to standard general
relativity in what concerns dynamical equations. The main difference
between both formulations is the existence, in the framework of
Teleparallel gravity, of a true gravitational energy-momentum vector
that is independent of coordinate transformations and sensible to
the change of reference frame. It replaces the Riemannian curvature
by the torsion in a tetrad formulation of Weitzenb\"{o}ck (or
Cartan) space-time \cite{Shirafuji,JWMaluf}, an approach originally
considered by Einstein himself in 1930~\cite{einstein}.

The tetrad field and metric tensor are related by $g^{\mu\nu}=e^{a\mu}e_{a}\,^{\nu}$. The familiar theory of general relativity deals with the Christoffel symbols ${}^0\Gamma_{\mu \lambda\nu}$, as the connection of space-time. On the other hand, TEGR is formulated in terms of Cartan
connection~\cite{Cartan},
$\Gamma_{\mu\lambda\nu}=e^{a}\,_{\mu}\partial_{\lambda}e_{a\nu}$. The geometric framework of both theories is related by means the following identity

\begin{equation}
\Gamma_{\mu \lambda\nu}= {}^0\Gamma_{\mu \lambda\nu}+ K_{\mu
\lambda\nu}\,, \label{2}
\end{equation}
where $K_{\mu \lambda\nu}$ is given by

\begin{eqnarray}
K_{\mu\lambda\nu}&=&\frac{1}{2}(T_{\lambda\mu\nu}+T_{\nu\lambda\mu}+T_{\mu\lambda\nu})\,.\label{3}
\end{eqnarray}
$K_{\mu\lambda\nu}$ is the contortion tensor defined in terms of the
torsion tensor constructed from the  Cartan connection. The torsion
tensor is $T_{\mu \lambda\nu}=e_{a\mu}T^{a}\,_{\lambda\nu}$, with

\begin{equation}
T^{a}\,_{\lambda\nu}=\partial_{\lambda} e^{a}\,_{\nu}-\partial_{\nu}
e^{a}\,_{\lambda}\,. \label{4}
\end{equation}

The curvature tensor obtained from $\Gamma_{\mu \lambda\nu}$ is
identically zero. From the identity (\ref{2}) we have

\begin{equation}
eR(e)\equiv -e({1\over 4}T^{abc}T_{abc}+{1\over
2}T^{abc}T_{bac}-T^aT_a)+2\partial_\mu(eT^\mu)\,,\label{eq5}
\end{equation}
where $R(e)$ is the scalar curvature of a Riemannian manifold in
terms of the tetrad field and $T^\mu=T^b\,_b\,^\mu$.

The Teleparallel Lagrangian density can be defined from (\ref{eq5}) and it reads

\begin{eqnarray}
\mathfrak{L}(e_{a\mu})&=& -\kappa\,e\,({1\over 4}T^{abc}T_{abc}+
{1\over 2} T^{abc}T_{bac} -T^aT_a) -\mathfrak{L}_M\nonumber \\
&\equiv&-\kappa\,e \Sigma^{abc}T_{abc} -\mathfrak{L}_M\;, \label{6}
\end{eqnarray}
where $\kappa=1/(16 \pi)$, $\mathfrak{L}_M$ is the Lagrangian
density of matter fields and $\Sigma^{abc}$ is given by

\begin{equation}
\Sigma^{abc}={1\over 4} (T^{abc}+T^{bac}-T^{cab}) +{1\over 2}(
\eta^{ac}T^b-\eta^{ab}T^c)\;, \label{7}
\end{equation}
with $T^a=e^a\,_\mu T^\mu$. It is important to note that the total divergence has been dropped once it does not contribute to the field equations. Thus both theories
share the same field equations and hence are dynamically equivalent to each other, since there is an equivalence between their Lagrangian densities.

The most general Lagrangian density in the realm of teleparallelism is given by

\begin{equation}
\mathfrak{L}=-e\,f(T)-\mathfrak{L}_M\,,\label{7.1}
\end{equation}
where $T=\Sigma^{abc}T_{abc}$. Performing a variational derivative of the above Lagrangian density with
respect to $e_{a \gamma}$, the dynamical variables of the
system, the field equations are

\begin{equation}
f'(T)\left[\partial_\nu \left(e\,\Sigma^{a\gamma\nu}\right)-e\,\Sigma^{bc\gamma}T_{bc}\,^a\right]-e\,\Sigma^{a\lambda\gamma}\left(\partial_\lambda T\right)f''(T)+\frac{1}{4}e\,e^{a\gamma}f(T)= {1\over {4\kappa}}\,e\,T^{a\gamma}\,,
\label{8}
\end{equation}
where $T^{a\mu}=e^{a}\,_{\lambda}T^{\mu
\lambda}=\frac{1}{e}\frac{\delta {\mathcal{L}}_{M}}{\delta e_{a\mu
}}$ is the energy-momentum tensor of matter fields. The prime in $f(T)$ means a
derivative with respect to $T$. The field equations can be rewritten as

\begin{equation}
\partial_\nu\left(e\Sigma^{a\lambda\nu}\,f'(T)\right)={1\over {4\kappa}}
e\, e^a\,_\mu( t^{\lambda \mu} + T^{\lambda \mu})\;, \label{10}
\end{equation}
where $t^{\lambda\mu}$ is defined by

\begin{equation}
t^{\lambda \mu}=\kappa\left[4\,f'(T)\,\Sigma^{bc\lambda}T_{bc}\,^\mu- g^{\lambda
\mu}\,f(T)\right]\,. \label{11}
\end{equation}
Since $\Sigma^{a\lambda\nu}$ is skew-symmetric in the last two
indices, it follows that

\begin{equation}
\partial_\lambda\partial_\nu\left(e\Sigma^{a\lambda\nu}\,f'(T)\right)\equiv0\,.\label{12}
\end{equation}
Thus we get

\begin{equation}
\partial_\lambda(et^{a\lambda}+eT^{a\lambda})=0\,\label{13}
\end{equation}
which yields the continuity equation

$$
{d\over {dt}} \int_V d^3x\,e\,e^a\,_\mu (t^{0\mu} +T^{0\mu})
=-\oint_S dS_j\, \left[e\,e^a\,_\mu (t^{j\mu} +T^{j\mu})\right] \,.
$$
It should be noted that the above expression works as a conservation
law for the sum of the energy-momentum tensor of matter fields and the quantity $t^{\lambda \mu}$. Thus $t^{\lambda \mu}$ is
interpreted as the energy-momentum tensor of the gravitational field
in the context of $f(T)$ theories, being more general than (and
slightly different from) the usual quantity in TEGR~\cite{maluf2},
\cite{PhysRevLett.84.4533}. Therefore, one can write the total
energy-momentum contained in a three-dimensional volume $V$ of space
as

\begin{equation}
P^a = \int_V d^3x \,e\,e^a\,_\mu(t^{0\mu}+ T^{0\mu})\,, \label{14}
\end{equation}
or using the field equations we have
\begin{equation}
P^a =4k\, \int_V d^3x \,\partial_\nu\left(e\,\Sigma^{a0\nu}\,f'(T)\right)\,. \label{14.1}
\end{equation}
It is worth noting that the above expression is invariant under
coordinate transformation and transforms like a vector under global
Lorentz transformations. Such features are desirable and expected
for a true energy-momentum vector, they are shared by other theories
such as TEGR and special relativity. However at this point it is
important to stress that the above expression is not invariant under
local Lorentz transformations due to the lack of such invariance in
the lagrangian density itself. In addition we point out that it is
impossible to get a local Lorentz invariance from a quantity defined
by integration such as eq. (\ref{14.1}) since it cannot depend on
coordinates, which precludes a transformation like
$P^a=\lambda^a\,_b(x) P^b$, where $\lambda^a\,_b(x)$ represents
local Lorentz transformations. From expression (\ref{eq5}) we see
that the scalar of curvature shares the same properties with the
scalar of torsion which reflect on the field equations. On the other
hand the picture is entirely different when one is dealing with
$f(T)$ theories since the total divergence is no longer in the
lagrangian density. As a consequence $f(T)$ theories are not
equivalent to $f(R)$ theories and different tetrads, even when originating
the same metric tensor, can lead to different field equations. Equation (\ref{14.1}) is unique in the sense that  for each tetrad
that is a solution of the field equations we have a well defined
gravitational energy-momentum vector.

Therefore the question on how to choose a tetrad field arises. This problem is addressed in ref. \refcite{Tamanini:2012hg} for spherical symmetry and FRW metric. There the authors define a ``good'' tetrad field by the following specifications: i) The solution of its field equations should reduce to a general relativity solution in the limit $f(T)\rightarrow T$; ii) The functional form of $f(T)$ should not be constrained by the field equations.

\section{Application in Cosmological Scales: The energy of the Universe}\label{frw}
\noindent

The cosmological principle asserts that the large-scale structure of
the Universe reveals homogeneity and isotropy~\cite{weinberg}. The
most general form of a line element that preserves such features may
be written as~\cite{Dinverno},

\begin{equation}
ds^2 = -dt^2 + a^2(t)\biggl[{dr^2 \over\left( 1-k'r^2\right)} + r^2(d\theta^2 +
\sin^2\theta d\phi^2) \biggr]\,,\label{5}
\end{equation}
where $a(t) = S(t)/|K|^{1/2}$ if $K \ne 0$ and $a(t) = S(t)$ if $K =
0$. $S(t)$ is the scale factor and $K$ is the constant curvature of
space. Here $K=|K|k'$ where $k'$ assumes the values $+1, 0, -1$ which
correspond to a space of constant positive curvature,
a flat space or a space of constant negative curvature,
respectively.

The simplest choice form for the tetrad field is diagonal, however it is not a good choice for it since such tetrad constrains the functional form of $f(T)$. It forces, by means the field equations, $f(T)$ to reproduce TEGR. It is shown in \refcite{Tamanini:2012hg} that, among all the possible tetrads, the ``good'' choice for $k=1$ would be

\begin{multline}
  {e_\mu}^a =
  \left(
  \begin{array}{cc}
    1 & 0  \\
    0 & a \cos\phi  \sin\theta /\sqrt{1-r^2} \\
    0 & r\, a \left(\sqrt{1-r^2} \cos \theta  \cos \phi -r \sin\phi \right)  \\
    0 & r\, a \sin\theta  \left(-r \cos\theta  \cos \phi -\sqrt{1- r^2} \sin\phi\right)
  \end{array}\right.\\
  \left.
  \begin{array}{cc}
    0 & 0 \\
    a \sin \theta  \sin \phi /\sqrt{1- r^2} & a \cos \theta /\sqrt{1- r^2} \\
    r\, a \left(r \cos\phi +\sqrt{1- r^2} \cos\theta  \sin\phi
    \right) & -r \sqrt{1- r^2}\, a \sin\theta \\
    r\, a \sin \theta  \left(\sqrt{1- r^2} \cos \phi -
    r \cos \theta  \sin \phi \right) & r^2\, a \sin ^2\theta
  \end{array}
  \right) \,,
  \label{5.1}
\end{multline}
where the above tetrad is presented with inverted lines and columns for the sake of adjustment.

The non-vanishing components of the torsion tensor are

\begin{eqnarray}
T_{101}&=&\left(\frac{a}{1-r^2}\right)\,\frac{\partial a}{\partial t}\,,\nonumber\\
T_{123}&=&\left(\frac{2a^2r^2}{\sqrt{1-r^2}}\right)\sin\theta\,,\nonumber\\
T_{202}&=&ar^2\,\left(\frac{\partial a}{\partial t}\right)\,,\nonumber\\
T_{213}&=&-\left(\frac{2a^2r^2}{\sqrt{1-r^2}}\right)\sin\theta\,,\nonumber\\
T_{303}&=&ar^2\,\left(\frac{\partial a}{\partial t}\right)\,\sin^2\theta\,,\nonumber\\
T_{312}&=&\left(\frac{2a^2r^2}{\sqrt{1-r^2}}\right)\sin\theta\,.
\end{eqnarray}

Then after some algebraic manipulations it is possible to obtain the scalar of torsion which reads

$$T=6\,\left[{\frac {-1+ \left( {\frac {da \left( t \right)}{dt}}  \right)
^{2}}{ \left( a\left( t \right)  \right) ^{2}}}\right]\,.
$$
The field equations, which establish the temporal evolution of $a\left( t \right)$, for $k=1$ are given by

\begin{eqnarray}
12\left(H\right)^2\,f^{\prime}(T)+f(T)&=&16\pi\rho\,,\\
\dot{H}(12H^2f^{\prime\prime}(T)-f^{\prime}(T))+\left(\frac{12H^2f^{\prime\prime}(T)+f^{\prime}(T)}{a^2}\right)&=&4\pi(\rho+ p)\,,
\end{eqnarray}
where $\rho$ and $p$ are respectively the energy density and the pressure of the cosmological perfect fluid. $H=\frac{\dot{a}}{a}$ is the Hubble constant.

Although the tetrad (\ref{5.1}) represent a good choice, it yields $$\Sigma^{(0)0i}=0\,,$$ which gives a vanishing expression for the gravitational energy for all possible functional form of $f(T)$, since
$$
P^{(0)} =4k\, \int_V d^3x \,\partial_i\left(e\,\Sigma^{(0)0i}\,f'(T)\right)\,.
$$
In addition, the procedure proposed in \refcite{Tamanini:2012hg} does not allow a good choice for the tetrad field when $k=-1$. It is important to point out that we do expect a vanishing gravitational energy when $k=0$, since we would be dealing with a flat Universe.

Other attempts have been made to construct good tetrads in the cosmological context, for instance one may see the tetrad field presented in ref.~\refcite{Ferraro:2011zb}. It represents the same tetrad above in different coordinates, as a consequence it will lead to the same vanishing gravitational energy. This means that in the context of $f(T)$ theories the gravitational energy is not due to the curvature of the spacetime, since it yields a vanishing one even in the presence of a spacetime with constant (and positive) curvature such as a closed Universe.

\section{Conclusion}
\noindent

In this paper we have obtained a general expression for the
gravitational energy-momentum in the realm of teleparallel gravity
when dynamics are governed by a general lagrangian as a function
of the scalar $T=\Sigma^{abc}T_{abc}$, and thus called $f(T)$
theories (in analogy to the $f(R)$ theories). Such an expression has
never appeared in the literature. We then applied our expression to
the FRW metric which was established taking into account the
principle of isotropy and homogeneity of the large scale of the
Universe. We show that the tetrad field that does not constrain the
functional form of the lagrangian density yields a vanishing energy.
We apply our expression for a tetrad obtained in the literature then
we show that the gravitational energy vanishes independent of the
functional form of $f(T)$, which may vary from the Born-Infeld model to
simpler ones that may represent small deviations of the TEGR
lagrangian density such as $f(T)=T+\frac{1}{2}\lambda T^2$. Thus we
conclude that in the context of $f(T)$ theories, for FRW spacetime,
the dynamics of a particle is due to the matter fields, which means
that the gravitational energy plays no role in such a system.


\begin{thebibliography}{10}

\bibitem{Felice}
Antonio~De Felice and Shinji Tsujikawa.
\newblock f(r) theories.
\newblock {\em Living Reviews in Relativity}, 13(3), 2010.

\bibitem{S.Perlmutter}
S.~Perlmutter, G.~Aldering, G.~Goldhaber, R.~A. Knop, P.~Nugent, P.~G. Castro,
  S.~Deustua, S.~Fabbro, A.~Goobar, D.~E. Groom, I.~M. Hook, A.~G. Kim, M.~Y.
  Kim, J.~C. Lee, N.~J. Nunes, R.~Pain, C.~R. Pennypacker, R.~Quimby,
  C.~Lidman, R.~S. Ellis, M.~Irwin, R.~G. McMahon, P.~Ruiz-Lapuente, N.~Walton,
  B.~Schaefer, B.~J. Boyle, A.~V. Filippenko, T.~Matheson, A.~S. Fruchter,
  N.~Panagia, H.~J.~M. Newberg, W.~J. Couch, and The Supernova~Cosmology
  Project.
\newblock Measurements of Ω and Λ from 42 high-redshift supernovae.
\newblock {\em The Astrophysical Journal}, 517(2):565, 1999.

\bibitem{1538-3881-116-3-1009}
Adam~G. Riess, Alexei~V. Filippenko, Peter Challis, Alejandro Clocchiatti, Alan
  Diercks, Peter~M. Garnavich, Ron~L. Gilliland, Craig~J. Hogan, Saurabh Jha,
  Robert~P. Kirshner, B.~Leibundgut, M.~M. Phillips, David Reiss, Brian~P.
  Schmidt, Robert~A. Schommer, R.~Chris Smith, J.~Spyromilio, Christopher
  Stubbs, Nicholas~B. Suntzeff, and John Tonry.
\newblock Observational evidence from supernovae for an accelerating universe
  and a cosmological constant.
\newblock {\em The Astronomical Journal}, 116(3):1009, 1998.

\bibitem{D.N.Spergel}
D.~N. Spergel, L.~Verde, H.~V. Peiris, E.~Komatsu, M.~R. Nolta, C.~L. Bennett,
  M.~Halpern, G.~Hinshaw, N.~Jarosik, A.~Kogut, M.~Limon, S.~S. Meyer, L.~Page,
  G.~S. Tucker, J.~L. Weiland, E.~Wollack, and E.~L. Wright.
\newblock First-year wilkinson microwave anisotropy probe (wmap) observations:
  Determination of cosmological parameters.
\newblock {\em The Astrophysical Journal Supplement Series}, 148(1):175, 2003.

\bibitem{0067-0049-170-2-377}
D.~N. Spergel, R.~Bean, O.~Doré, M.~R. Nolta, C.~L. Bennett, J.~Dunkley,
  G.~Hinshaw, N.~Jarosik, E.~Komatsu, L.~Page, H.~V. Peiris, L.~Verde,
  M.~Halpern, R.~S. Hill, A.~Kogut, M.~Limon, S.~S. Meyer, N.~Odegard, G.~S.
  Tucker, J.~L. Weiland, E.~Wollack, and E.~L. Wright.
\newblock Three-year wilkinson microwave anisotropy probe (wmap) observations:
  Implications for cosmology.
\newblock {\em The Astrophysical Journal Supplement Series}, 170(2):377, 2007.

\bibitem{M.Tegmark}
Max Tegmark, Michael~A. Strauss, Michael~R. Blanton, Kevork Abazajian, Scott
  Dodelson, Havard Sandvik, Xiaomin Wang, David~H. Weinberg, Idit Zehavi,
  Neta~A. Bahcall, Fiona Hoyle, David Schlegel, Roman Scoccimarro, Michael~S.
  Vogeley, Andreas Berlind, Tam\'as Budavari, Andrew Connolly, Daniel~J.
  Eisenstein, Douglas Finkbeiner, Joshua~A. Frieman, James~E. Gunn, Lam Hui,
  Bhuvnesh Jain, David Johnston, Stephen Kent, Huan Lin, Reiko Nakajima,
  Robert~C. Nichol, Jeremiah~P. Ostriker, Adrian Pope, Ryan Scranton,
  Uro\ifmmode \check{s}\else~\v{s}\fi{} Seljak, Ravi~K. Sheth, Albert Stebbins,
  Alexander~S. Szalay, Istv\'an Szapudi, Yongzhong Xu, James Annis,
  J.~Brinkmann, Scott Burles, Francisco~J. Castander, Istvan Csabai, Jon
  Loveday, Mamoru Doi, Masataka Fukugita, Bruce Gillespie, Greg Hennessy,
  David~W. Hogg, \ifmmode \check{Z}\else~\v{Z}\fi{}eljko
  Ivezi\ifmmode~\acute{c}\else \'{c}\fi{}, Gillian~R. Knapp, Don~Q. Lamb,
  Brian~C. Lee, Robert~H. Lupton, Timothy~A. McKay, Peter Kunszt, Jeffrey~A.
  Munn, Liam O'Connell, John Peoples, Jeffrey~R. Pier, Michael Richmond,
  Constance Rockosi, Donald~P. Schneider, Christopher Stoughton, Douglas~L.
  Tucker, Daniel~E. Vanden~Berk, Brian Yanny, and Donald~G. York.
\newblock Cosmological parameters from sdss and wmap.
\newblock {\em Phys. Rev. D}, 69:103501, May 2004.

\bibitem{PhysRevD.71.103515}
Uro\ifmmode \check{s}\else~\v{s}\fi{} Seljak, Alexey Makarov, Patrick McDonald,
  Scott~F. Anderson, Neta~A. Bahcall, J.~Brinkmann, Scott Burles, Renyue Cen,
  Mamoru Doi, James~E. Gunn, \ifmmode \check{Z}\else~\v{Z}\fi{}eljko
  Ivezi\ifmmode~\acute{c}\else \'{c}\fi{}, Stephen Kent, Jon Loveday, Robert~H.
  Lupton, Jeffrey~A. Munn, Robert~C. Nichol, Jeremiah~P. Ostriker, David~J.
  Schlegel, Donald~P. Schneider, Max Tegmark, Daniel E.~Vanden Berk, David~H.
  Weinberg, and Donald~G. York.
\newblock Cosmological parameter analysis including sdss $\alpha$ forest and
  galaxy bias: Constraints on the primordial spectrum of fluctuations, neutrino
  mass, and dark energy.
\newblock {\em Phys. Rev. D}, 71:103515, May 2005.

\bibitem{D.J.Eisenstein}
Daniel~J. Eisenstein, Idit Zehavi, David~W. Hogg, Roman Scoccimarro, Michael~R.
  Blanton, Robert~C. Nichol, Ryan Scranton, Hee-Jong Seo, Max Tegmark, Zheng
  Zheng, Scott~F. Anderson, Jim Annis, Neta Bahcall, Jon Brinkmann, Scott
  Burles, Francisco~J. Castander, Andrew Connolly, Istvan Csabai, Mamoru Doi,
  Masataka Fukugita, Joshua~A. Frieman, Karl Glazebrook, James~E. Gunn, John~S.
  Hendry, Gregory Hennessy, Zeljko Ivezić, Stephen Kent, Gillian~R. Knapp,
  Huan Lin, Yeong-Shang Loh, Robert~H. Lupton, Bruce Margon, Timothy~A. McKay,
  Avery Meiksin, Jeffery~A. Munn, Adrian Pope, Michael~W. Richmond, David
  Schlegel, Donald~P. Schneider, Kazuhiro Shimasaku, Christopher Stoughton,
  Michael~A. Strauss, Mark SubbaRao, Alexander~S. Szalay, István Szapudi,
  Douglas~L. Tucker, Brian Yanny, and Donald~G. York.
\newblock Detection of the baryon acoustic peak in the large-scale correlation
  function of sdss luminous red galaxies.
\newblock {\em The Astrophysical Journal}, 633(2):560, 2005.

\bibitem{ferraro}
Rafael Ferraro and Franco Fiorini.
\newblock Modified teleparallel gravity: Inflation without an inflaton.
\newblock {\em Phys. Rev. D}, 75:084031, Apr 2007.

\bibitem{Bengochea}
Gabriel~R. Bengochea and Rafael Ferraro.
\newblock Dark torsion as the cosmic speed-up.
\newblock {\em Phys. Rev. D}, 79:124019, Jun 2009.

\bibitem{Linder}
Eric~V. Linder.
\newblock Einstein's other gravity and the acceleration of the universe.
\newblock {\em Phys. Rev. D}, 81:127301, Jun 2010.

\bibitem{Ulhoa:2010wv}
S.C. Ulhoa, J.F da~Rocha~Neto, and J.W. Maluf.
\newblock {The Gravitational Energy Problem for Cosmological Models in
  Teleparallel Gravity}.
\newblock {\em Int.J.Mod.Phys.}, D19:1925--1935, 2010.

\bibitem{Shirafuji}
Kenji Hayashi and Takeshi Shirafuji.
\newblock New general relativity.
\newblock {\em Phys. Rev. D}, 19:3524--3553, Jun 1979.

\bibitem{JWMaluf}
Jos\'{e}~W. Maluf.
\newblock Hamiltonian formulation of the teleparallel description of general
  relativity.
\newblock {\em Journal of Mathematical Physics}, 35(1):335--343, 1994.

\bibitem{einstein}
A.~Einstein.
\newblock Unified field theory based on riemannian metrics and distant
  parallelism.
\newblock {\em Math. Annal.}, 102:685--697, 1930.

\bibitem{Cartan}
E.~{Cartan}.
\newblock {On a Generalization of the Notion of Reimann Curvature and Spaces
  with Torsion}.
\newblock In P.~G. {Bergmann} and V.~{de Sabbata}, editors, {\em NATO ASIB
  Proc. 58: Cosmology and Gravitation: Spin, Torsion, Rotation, and
  Supergravity}, pages 489--491, 1980.

\bibitem{maluf2}
J.~W. Maluf.
\newblock {The gravitational energy-momentum tensor and the gravitational
  pressure}.
\newblock {\em Annalen Phys.}, 14:723--732, 2005.

\bibitem{PhysRevLett.84.4533}
V.~C. de~Andrade, L.~C.~T. Guillen, and J.~G. Pereira.
\newblock Gravitational energy-momentum density in teleparallel gravity.
\newblock {\em Phys. Rev. Lett.}, 84:4533--4536, May 2000.

\bibitem{Tamanini:2012hg}
Nicola Tamanini and Christian~G. Boehmer.
\newblock {Good and bad tetrads in f(T) gravity}.
\newblock {\em Phys.Rev.}, D86:044009, 2012.

\bibitem{weinberg}
S.~Weinberg.
\newblock {\em Gravitation and Cosmology: Principles and Applications of the
  General Theory of Relativity}.
\newblock John and Wiley \& Sons, Inc., 1972.

\bibitem{Dinverno}
Ray d'Inverno.
\newblock {\em Introducing Einstein's Relativity}.
\newblock Clarendon Press, Oxford, 4th edition, 1996.

\bibitem{Ferraro:2011zb}
Rafael Ferraro and Franco Fiorini.
\newblock {Cosmological frames for theories with absolute parallelism}.
\newblock {\em Int.J.Mod.Phys.Conf.Ser.}, 3:227--237, 2011.

\end{thebibliography}

\end{document}